\documentclass[twocolumn,aps,amsmath,amssymb,showpacs,floatfix]{revtex4}
\usepackage{graphicx}% Include figure files 
\usepackage{dcolumn}% Align table columns on decimal point 
\usepackage{bm}% bold math
\usepackage{hyperref} 
\usepackage{latexsym}
\usepackage{float}
\def\av#1{\langle#1\rangle}
\begin{document}
\title{First Passage Properties of the Erd\H{o}s-Renyi Random Graph}   
\author{V. Sood}
\email{vsood@bu.edu}
\author{S.~Redner}
\email{redner@bu.edu}
\altaffiliation{First two authors on leave of absence from Department of Physics, Boston University}
\affiliation{Theory Division  and Center for Nonlinear Studies, Los Alamos National Laboratory, Los
Alamos, New Mexico 87545}
\author{D. ben-Avraham}
\email {qd00@clarkson.edu}
\affiliation{Department of Physics, Clarkson University, Potsdam NY 13699, USA}
\begin{abstract} 
  We study the mean time for a random walk to traverse between two arbitrary
  sites of the Erd\H{o}s-Renyi random graph.  We develop an effective medium
  approximation that predicts that the mean first-passage time between pairs
  of nodes, as well as all moments of this first-passage time, are
  insensitive to the fraction $p$ of occupied links.  This prediction
  qualitatively agrees with numerical simulations away from the percolation
  threshold.  Near the percolation threshold, the statistically meaningful
  quantity is the mean transit rate, namely, the inverse of the first-passage
  time.  This rate varies non-monotonically with $p$ near the percolation
  transition.  Much of this behavior can be understood by simple heuristic
  arguments.
\end{abstract}

\maketitle

\date{\today}

\section{Introduction}

In this article, we study a basic first-passage characteristic of random
walks on Erd\H{o}s-Renyi (ER) random graphs \cite{ER}, namely, the mean time
for a random walk to traverse between two arbitrary sites on the graph.  The
ER random graph is constructed by taking $N$ sites and introducing a bond
between each pair of sites with probability $p$.  When $p=1$, all possible
links exist and this construction gives the complete graph, where each site
is connected to all the other $N-1$ sites in the graph.  As $p$ decreases,
the random graph undergoes a percolation transition at $p=p_c=1/N$
\cite{B,JLR} that shares many common features with percolation on regular
lattices.  Another geometrical feature that is relevant for our study of
first-passage characteristics is a second connectivity transition at
$p_1=\ln N/N$.  For $p>p_1$, all nodes belong to a single component (in the
limit $N\to\infty$), while for $p<p_1$ disjoint clusters can exist
\cite{B,JLR}.

Much effort has been devoted to determining basic properties of random walks
in disordered or heterogeneous environments, \cite{L,W,AF,AH00}, such as the
ER random graph, as well as small-world \cite{AKS} and scale-free networks
\cite{NR}.  On the random graph, there has been considerable work in
determining how basic time scales of the random walk depend on the size of
the graph.  These include the mixing time -- the time scale that determines
how the probability distribution approaches its limiting behavior
\cite{BR,M,AKS}, the cover time -- the time for a random walk to visit all
sites of the graph \cite{CF}, and the first-passage time -- the time for a
random walk to traverse between two specified points for the first time
\cite{NR} or to return to its starting point for the first time \cite{MK}.

The goal of this work is to understand by simple physical arguments how a
basic first-passage property of random walks on random graphs depends on the
concentration of bonds in the graph.  We will begin by studying the mean
first-passage, or transit, time $T_{ab}$ for a random walk to reach an
arbitrary site $b$ on the graph when starting from another arbitrary site
$a$.  In general, this transit time $T_{ab}$ does not necessarily equal
$T_{ba}$ and it is sometimes convenient to consider the mean commute time
$K_{ab} \equiv T_{ab} + T_{ba}$ to avoid the asymmetry in $T_{ab}$.  However,
in averaging over all pairs of sites $a$ and $b$ to obtain statistically
meaningful quantities, the asymmetry is eliminated and
$\av{K_{ab}}=2\av{T_{ab}}$.

In practice, mean transit and commute times diverge when the graph consists
of more than one component (cluster), because a random walk that starts in
one component cannot access sites in different components.  We are thus led
to study the dependence of the mean commute rate ${\cal R}_{ab}\equiv
1/K_{ab}$ as a function of $p$ on the random graph.  This rate equals zero
for two sites in different components, so that its configurational average is
meaningful.  Focusing on the commute rate is analogous to considering mean
conductance of a random conductor-insulator mixture near the percolation
threshold, rather than the mean resistance.  The conductance is well-behaved
near the percolation threshold, while the resistance is divergent for all $p$
in any finite-size system.

In the next section, we construct an effective medium approximation that
predicts that the mean commute time is independent of $p$ for $p>p_1$.  We
also find that the transit time varies weakly with $p$ in the small-dilution
limit.  In Sec.~III, we present simulation results for the mean commute time
and the mean commute rate and find an unexpected non-monotonic behavior for
the latter quantity as a function of $p$ when $p\approx p_c$.  In Sec.~IV, we
outline the relation between the commute time on a network and the two-point
conductance on the same network when each link is a unit resistor.  We use
this electrical network connection to explain the non-monotonicity of the
commute rate, first for a tree structure that is a subset of the random graph
(Sec.~V), and then for the random graph itself (Sec.~VI).

\section{Analytic Approaches for the Transit Time}

\subsection{Effective Medium Approach}

We now develop an effective medium approximation for the commute time of a
discrete-time random walk on the random graph \cite{emt-rev}.  In a single
time step, a walk located at a site that is connected to $z$ other sites can
hop with probability $1/z$ to any of these neighbors.  To compute the mean
time for such a random walk to go between two arbitrary sites on any graph by
a sequence of nearest-neighbor hops, we use the underlying backward equation
\cite{vk,fpp}.  This equation relates the transit time from site $a$ to site
$b$ to the transit times from the neighboring sites of $a$ to site $b$ as
follows:
\begin{equation}
\label{tt}
T_{ab} = \sum_\Pi P_{\,\Pi} \,\,t_{\,\Pi}= \sum_i p_{a\to i}(\delta t+T_{ib}).
\end{equation}
The first sum is over all paths $\Pi$ from $a$ to $b$, $P_\Pi$ is the
probability for the random walk to take the path $\Pi$, and $t_\Pi$ is the
transit time from $a$ to $b$ along this path.  For each path, we then
decompose the full transit time into the time to go from $a$ to an
intermediate site $i$ after one step plus the time to go from $i$ to $b$.
Thus $p_{a\to i}= 1/z_a$ is the probability of hopping from $a$ to $i$ in a
single step, $z_a$ is the degree of $a$, and $\delta t$ is the time for each
step of the random walk.  Without loss of generality, we take $\delta t = 1$.

\begin{figure}[ht]
\vspace*{0.cm}
\includegraphics*[width=0.2\textwidth]{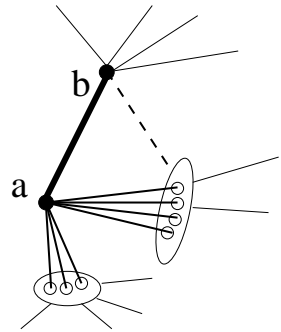}
\caption{Schematic decomposition of a random graph with starting site $a$ 
  and target site $b$.  The direct link is shown as the thick solid line.
  After one step via an indirect path (to the sites in of the ovals), either
  $b$ can be reached directly with probability $p$ (dashed line), while with
  probability $q=1-p$ there is still no direct connection to $b$.}
\label{EMT}
\end{figure}

Let us now construct an effective-medium approximation for the average
transit time on the random graph, under the assumption that the graph is
connected.  This condition implicitly restricts the validity of our approach
to the range $p>p_1$, where all nodes belong to a single component.  A
schematic representation of a random graph, to illustrate our approach, is
shown in Fig.~\ref{EMT}.  Between two sites $a$ and $b$ on the graph, a
direct link to $b$ may exist (thick line) with probability $p$.  If there is
no such direct link, then an indirect path must be followed.  After a single
step on this indirect path (medium lines), there may be a direct link to $b$
with probability $p$ (dashed), or no direct link with probability $q=1-p$.

Let us denote by $\tau$ the mean transit time to go from $a$ to $b$ under the
assumption that a direct link exists, and $\tau^\prime$ the transit time from
$a$ to $b$ in the absence of a direct link.  Then from Eq.~(\ref{tt}) and
following an effective-medium assumption, $\tau$ obeys the recursion formula
\begin{equation}
\label{tMfappx}
\tau = \frac{1}{(N-1)p}+\left[1 - \frac{1}{(N-1)p}\right]\left[\,p(1+\tau)
+q(1+\tau^\prime)\,\right].
\end{equation}
The first term accounts for the walk that goes directly from $a$ to $b$.
This contribution corresponds, in Eq.~(\ref{tt}), to the case where the
intermediate site $i$ coincides with $b$.  Since $(N-1)p$ links emanate from
$a$ on average, then according to the effective-medium approximation, the
probability that a random walk steps along the direct connection is just
$1/(N-1)p$.  The second set of terms accounts for those walks in which the
first step goes to an intermediate site $i$ rather than hitting $b$ directly.
In this case, we again apply an effective-medium approximation and posit that
after one step of the walk, a direct connection from $i$ to $b$ exists with
probability $p$, or no direct connection exists with probability $1-p$
(Fig.~\ref{EMT}).

To close this equation, we need an expression for $\tau^\prime$, the
first-passage time in the absence of a direct connection to $b$.  Applying
the same effective-medium approximation as that used in Eq.~(\ref{tMfappx}),
we assume that after the first step of the walk, the terminal site $b$ is
directly reachable with probability $p$, while $b$ is not directly reachable
with probability $q$.  Thus $\tau^\prime $ obeys
\begin{equation}
\label{tprime-rg}
\tau'=p(1+\tau)+q(1+\tau^\prime).
\end{equation}

Solving Eqs.~(\ref{tMfappx}) \& (\ref{tprime-rg}) gives $\tau \!= \!N \!-
\!\frac{1}{p}$ and $\tau^\prime \!=\!N$.  Finally, we average the transit
time over all pairs of terminal points and over all graph configurations.
Again in the spirit of an effective medium approximation, this average is
simply
\begin{equation}
\label{tav}
\langle T\rangle\equiv\langle T_{ab}\rangle =p\tau +(1-p)\tau^\prime = N-1.
\end{equation}
Surprisingly, $\langle T\rangle$ is independent of $p$.  Thus according to
the effective medium approach, the complete graph solution, $\langle
T\rangle=N-1$, holds for all $p$.

The backward equation for the mean transit time can be extended to any positive
integer moment of the transit time.  Consider, for example, the mean-square
transit time.  As in the case of the mean time, the governing equation can
formally be written as
\begin{equation}
\label{t2-def}
T_{ab}^2=\sum_{\,\Pi}\, P_{\,\Pi} \,\,t_\Pi^2,
\end{equation}
For each path, we follow Eq.~(\ref{tt}) and again write the transit time
$t_{\,\Pi}$ as $1+t_{\,\Pi'}$, namely, the sum of the time for the first step
and the time for the remainder of the path.  Thus
\begin{eqnarray}
\label{decomp}
T_{ab}^2  &=& \sum_{\,\Pi} P_{\,\Pi} (1+t_{\,\Pi'})^2,\nonumber \\
   &=& \sum_{\,\Pi} P_{\,\Pi} (1+2t_{\,\Pi'}+ t_{\,\Pi'}^2),\nonumber \\
   &=& \sum_i p_{a\to i} (1+2\tau_i+ \tau_i^2).
\end{eqnarray}
In going from the second to the last line of this equation, we use the fact
that $P_{\,\Pi}=\sum_i p_iP_{\,\Pi'}$, where $p_i$ is the probability of
hopping from the starting point to one of its nearest neighbors $i$, and
$P_{\,\Pi'}$ is the probability for the remainder of the path $\Pi'$ from $i$
to $b$.  In the last line, the quantities $\tau_i$ and $\tau_i^2$ are the
mean and mean-square times to reach $b$ when starting from $i$ and the sum is
over all neighbors $i$ of the starting point.  Strictly speaking, we should
write $\langle\tau\rangle$ and $\langle\tau^2\rangle$ for these moments, so
that it is obvious that $\langle\tau^2\rangle \ne \langle\tau\rangle^2$.  In
the following, we drop these angle brackets because the linear and quadratic
powers of time always appear separately and there is no ambiguity about where
the angle brackets should appear.

The last line of Eq.~(\ref{decomp}) is now a backward equation for the second
moment of the first-passage time, in which the previously-determined first
moment is an input to this equation.  This construction for the mean-square
transit time can be generalized straightforwardly, albeit tediously, to any
positive integer moment of the first-passage time.  For the random graph, the
recursion formula for the mean-square transit time is, in close analogy with
Eqs.~(\ref{tMfappx}) and (\ref{tprime-rg}).
\begin{eqnarray*}
\label{t2-rg}
\tau^2&=&\frac{1}{(N-1)p}+
\left[1-\frac{1}{(N-1)p}\right]\times \nonumber \\
&&~~~~~~~\left[p(1+2\tau+\tau^2)+q(1+2\tau'+\tau'^2)\right],\nonumber
\\ 
\tau'^2&=&p(1+2\tau+\tau^2)+q(1+2\tau'+\tau'^2).
\end{eqnarray*}
Using our previously-derived results for the first moments,
$\tau=N-\frac{1}{p}$ and $\tau'=N$, these recursion formulae are easily
solved.  We then compute the configuration averaged mean-square transit time,
$\langle T^2\rangle \equiv p\tau^2+q\tau'^2$, and obtain $\langle T^2\rangle
=(2N-3)(N-1)$.  Thus again, the second moment is independent of $p$ and
equals the second moment of the transit time on the complete graph.

More generally, we show that the first-passage probability between any two
sites on a random graph, and thus all moments of the first-passage time, are
independent of $p$ in the effective-medium approximation.  As a preliminary,
we first compute the first-passage probability on the complete graph.  Let
$F(t)$ be the probability that a random walk hits the target site for the
first time at time $t$, and let $F(z)=\sum F(t)z^t$ be the corresponding
generating function.  For the complete graph, the generating function obeys
the recursion formula
\begin{eqnarray*}
F(z) = \frac{1}{N-1}\, z + \frac{N-2}{N-1}\, z \,F(z).
\end{eqnarray*}
This equation encodes the fact that after a single step (the factor $z$) the
walk hits the target site with probability $1/(N-1)$, while with probability
$(N-2)/(N-1)$ the walk hits another interior site of the graph, at which
point the first-passage process is renewed.  The solution to this equation is
\begin{eqnarray*}
F(z)=\frac{z}{N-1}\left[1-\left(\frac{N-2}{N-1}\right)\right]^{-1},
\end{eqnarray*}
from which 
\begin{eqnarray}
\label{F}
F(t)=\frac{1}{N-1}\left(\frac{N-2}{N-1}\right)^t.
\end{eqnarray}

Now consider the random graph with bond occupation probability $p$.  Let
${\cal F}(t)$ be the first-passage probability from $a$ to $b$ when a bond is
present between these two sites, ${\cal F}\,'(t)$ the first-passage probability
when this bond is absent, and let ${\cal F}(z)$ and ${\cal F}\,'(z)$ be the
respective generating functions.  In the spirit of our effective medium
approximation given in Eqs.~(\ref{tMfappx}) and (\ref{tprime-rg}), we now
have
\begin{eqnarray*}
{\cal F}(z) &\!=\!& \frac{1}{(N\!-\!1)p} z \!+\! \left[1\!-\!\frac{1}{(N\!-\!1)p}\right]
\left[ p z {\cal F}(z) + q z {\cal F}\,'(z) \right],\\ {\cal F}\,'(z) &\!=\!& p z
{\cal F}(z) + q z {\cal F}\,'(z).
\end{eqnarray*}
From these two equations, the average first-passage probability $\langle
F(z)\rangle = p {\cal F}(z) + q {\cal F}\,'(z)$ has the same form as the
first-passage probability for the complete graph (Eq.~(\ref{F})).  Hence all
moments of the transit time are independent of $p$ in the effective medium
approximation.

\subsection{Small Dilution Limit}

We may understand the exact dependence of the mean transit time in the limit
$p\to 1$ by considering configurations with a single missing bond.  There are
four distinct cases to consider: (i) missing link between $a$ and $b$ (1
configuration), (ii) missing link between $a$ and an interior point ($N-2$
configurations), (iii) missing link between an interior point and $b$ ($N-2$
configurations), and (iv) missing link between two interior points (all
remaining configurations).  Let us denote the mean transit times from $a$ to
$b$ for these 4 configurations by $t_1$, $t_2$, $t_3$, and $t_4$.  By
considering each case separately, we obtain the recursion formulae (cf.\
Eq.~(\ref{tt})):
\begin{eqnarray*}
\label{4t}
t_1&=&1+t_3 \nonumber \\
t_2&=&\frac{1}{N-2}+\frac{N-3}{N-2}(1+t_4)\nonumber \\
t_3&=&\frac{1}{N-1}+\frac{1}{N-1}(1+t_1)+ \frac{N-3}{N-1}(1+t_3)\nonumber\\
t_4&=&\frac{1}{N-1}+\frac{2}{N-1}(1+t_2)+ \frac{N-4}{N-1}(1+t_4),
\end{eqnarray*}
with solution
\begin{eqnarray}
\label{4t-soln}
t_1&=&N+1 \quad ~~~  t_2=N-1-\frac{3}{N} \nonumber \\
t_3&=&N \qquad ~~~~~ t_4=N-1-\frac{2}{N}.
\end{eqnarray}
Then, by averaging over the appropriate number of configurations for each
class, we obtain the mean transit time
\begin{equation}
\av{T}=\frac{N^3-2N^2+N+4}{N(N-1)}\approx N-1+\frac{4}{N^2}.
\end{equation}

To interpret this result, note that the absence of a single bond corresponds
to a bond concentration $p\approx1-2/N^2$.  At this value of $p$, the mean
transit time has the asymptotic behavior $\langle T\rangle/\langle
T(p\!=\!1)\rangle\sim (1+4/N^3)\sim [1+\sqrt{2}(1-p)^{3/2}]$.  The first
correction to $\langle T\rangle$ is thus of the order of $(1-p)^{3/2}$,
rather than linear in $(1-p)$, as one might naively expect.  This small first
correction to $\langle T\rangle$ near $p=1$ makes plausible the
effective-medium result that $\langle T\rangle$ is independent of $p$.

\section{Simulation Results}

To test the effective-medium prediction for the mean first-passage time, we
now turn to numerical simulations.  For very small systems ($N\leq 8$) , we
have obtained the exact first-passage time by averaging over all
configurations of random graphs, over all pairs of endpoints, and over all
random walks.  For the graph configuration average, each realization is
weighted by the factor $p^k q^{E-k}$, where $k$ is the number of occupied
links in the graph, $E=N(N-1)/2$ is the total number of possible links, and
$q=1-p$.  We then average over all pairs of endpoints directly.  By this
averaging, the mean transit time is simply one-half of the mean commute time.
Rather than averaging over individual walks directly, we solve exactly the
recursion formulae in Eq.~(\ref{tt}) for the transit times between all pairs
of points.

For larger systems, the exact enumeration of all graph configurations is
impractical.  Instead we average over a finite number of graph realizations
and endpoint pairs, but still performed the exact average over all random
walk trajectories by numerically solving Eq.~(\ref{tt}).  For efficiency, we
start our simulation with an empty graph, add bonds one at a time and then
update the commute times between all pairs of sites in the graph after each
bond addition.  Each graph is then weighted by $p^k q^{E-k}$ so that we can
obtain the commute time as a function of $p$.  We repeat this sequential
graph construction over many realizations.  The graphs that we obtain by this
sequential growth are the same as those obtained by a static construction in
which each bond is present with probability $p = {2M}/{N(N-1)}$ when $N$ is
large (see \cite{B} for different, but equivalent ways of constructing
random graphs).

\begin{figure}[ht]
\begin{center}
\includegraphics*[width=0.50\textwidth]{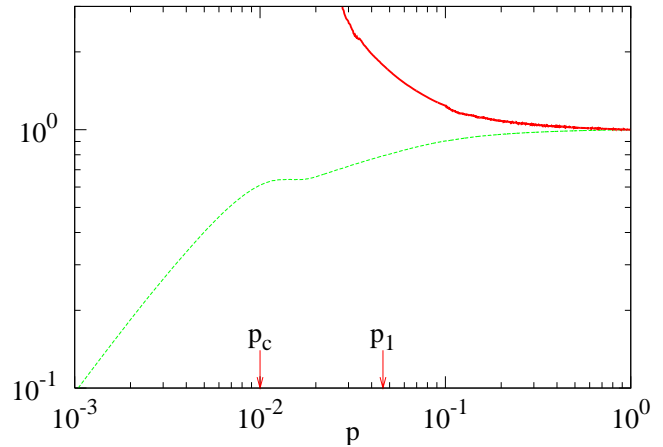}
\end{center}
\caption{Mean commute time (dashed) and mean commute rate (solid) versus bond
  occupation probability $p$ for a random graph of $N=100$ sites.  Both
  quantities are normalized to have the value 1 for the complete graph
  ($p=1$).  Averages over $10^3$ graph realizations were performed for each
  $p$.  Also shown are the locations of $p_c=0.01$ and $p_1\approx 0.046$.}
\label{CRCTFullRange}
\end{figure}

For the average commute time, we only include connected graphs in the
ensemble, while for the average rate, the ensemble consists of all graph
configurations.  This restriction plays a significant role only for $p<p_1$,
where the random graph normally consists of multiple components.  Typical
results for a graph of 100 sites are shown in Fig.~\ref{CRCTFullRange}.
Above the connectivity threshold $p_1=\ln N/N$, the average transit time
varies slowly with $p$, in agreement with our effective medium approach.  The
apparent singularity of the average commute time at a value $p<p_1$ stems
from finite size effects.

\begin{figure}[ht]
  \vspace*{0.cm}\includegraphics*[width=0.50\textwidth]{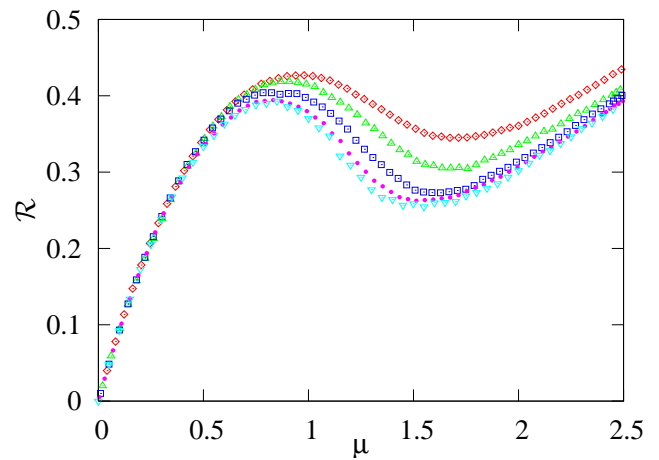}
\caption{Mean commute rate ${\mathcal R}$ on a random graph for $N=50$
  ($\diamond$), $N=100$ ($\bigtriangleup$), $200$ ($\square$), $400$
  ($\circ$), and 800 ($\bigtriangledown$) sites as a function of the average
  site degree $\mu = p(N-1)$.  These rates are normalized to one for the
  complete graph limit.  Averages over $10^3$ graphs were performed for each
  case.  }
\label{commuteRate-graph}
\end{figure}

The behavior of the mean commute rate is shown in Figs.~\ref{CRCTFullRange}
and ~\ref{commuteRate-graph}.  Unexpectedly, this rate is non-monotonic in
$p$ for $p\approx p_c={1}/{N}$, as shown in detail in
Fig.~\ref{commuteRate-graph}.  For this plot, we use the average degree, $\mu
= p(N-1)$ as the dependent variable, because it has the desirable feature
that the percolation transition occurs at the same value $\mu_c=1$ for all
$N$.  The fact that the non-monotonicity in the commute rate occurs near
$\mu=1$ suggests that this anomaly is connected with the percolation
transition of the random graph.

To understand this non-monotonicity, we first make a connection between the
commute rate and the conductance on the same network when each link is a unit
resistance, and then analyze the structure of the random graph in the
critical regime to evaluate the conductance.  This result will then be used
to infer the dependence of the commute rate on the mean degree $\mu$.

\section{Electrical Network Connection}

In principle, first-passage properties of random walks on a graph can be
obtained from the underlying Laplacian matrix of the graph \cite{L}.  The
eigenvalue spectrum of the Laplacian provides many time-dependent random walk
characteristics.  This matrix formulation also reveals deep analogies between
random walks on a graph and the electrical network problem on the same graph
in which each occupied link is a resistor of unit resistance (see
Ref.~\cite{DS1} for a nice exposition of these connections).

For example, the commute time $K_{ab}$ between $a$ and $b$ and the
conductance $G_{ab}$ between these same two sites are simply related by
$K_{ab} = 2M/G_{ab}$ \cite{B,L}.  Here $M$ is the number of bonds in the
cluster that contains both $a$ and $b$.  Equivalently, the mean commute rate
${\cal R}_{ab}=1/K_{ab}$ is given by
\begin{equation}
\label{Res-CT}
{\cal R}_{a b} = \frac{G_{a b}}{2M}.
\end{equation}
As we shall see, it is much easier to estimate the conductance rather than
the commute rate of a random graph by direct means.  We will then rely on
this connection between $G_{ab}$ and ${\cal R}_{a b}$ to determine the latter
quantity. 

For reasons of numerical convenience, we will often consider the following
sum of the rates
\begin{eqnarray}
\label{CR-cond}
{\cal R}_a \equiv 2\sum_{b\ne a} {\cal R}_{a b}=\frac{1}{M}\sum_b G_{a b}.
\end{eqnarray}  
We include the factor of 2 in the definition because ${\cal R}_a$ then equals
1 for the complete graph.  We may also sum freely over all sites $b$ in the
system Eq.~(\ref{CR-cond}) because $G_{ab}=0$ for any sites that are not in
the same cluster as $a$.  Finally, we obtain the average commute rate for the
graph by averaging over all initial sites $a$:
\begin{eqnarray}
\label{CR-graph}
{\cal R} \equiv \frac{1}{N}\sum_a {\cal R}_a.
\end{eqnarray}

\begin{figure}[ht]
\vspace*{0.cm}
\includegraphics*[width=0.500\textwidth]{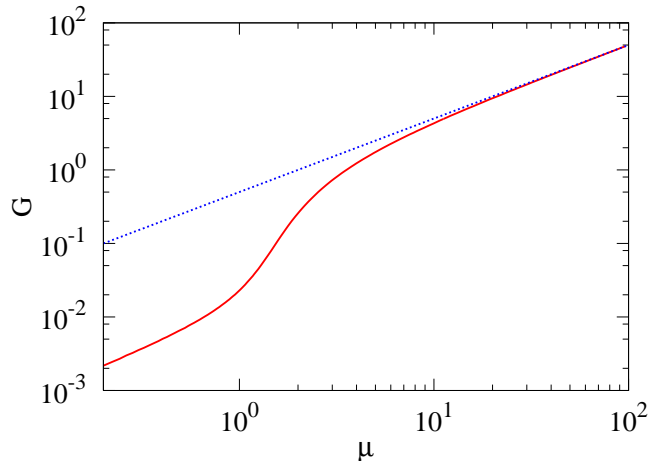}
\caption{Average two-point conductance (thick solid curve) on a random graph 
  with $N=100$ sites.  The dotted line corresponds to $G={\mu}/{2}$, the
  asymptotic large-$\mu$ form for the conductance.  
  }
\label{RG-cond}
\end{figure}

In the limit of large $\mu$ all the sites in the graph belong to the same
cluster and $M ={\mu N}/{2}$.  Thus the average commute rate becomes
\begin{equation}
\label{CR-DenseGraph}
{\cal R} = \frac{1}{N}\frac{2}{\mu N}\sum_{a b}G_{a b}=\frac{2}{\mu}\,\,G,
\end{equation}
where $G$ is the two-point conductance averaged over all pairs of graph
endpoints.  Thus, as we have discussed, first-passage times and two-point
conductances are intimately connected.

For the conductance itself, it is worth noting that this function behaves
anomalously near the connectivity transition.  Although the conductance must
increase monotonically with $\mu$ \cite{DS1}, the rate of increase changes
for $\mu$ in the critical range between 1 and $\ln N$ (Fig.~\ref{RG-cond}).
For large $\mu$, the conductance asymptotically approaches $G={\mu}/{2}$
(dashed line), a result that corresponds to the average commute rate
approaching ${\mathcal R}=1$, in agreement with the result of
Fig.~\ref{CRCTFullRange}.  

\section{Structure of the Random Graph}

To determine the behavior of the commute rate for general values of $\mu$, we
first need to resolve the structure of random graphs at these values of
$\mu$.  From this structural information, it is relatively easy to determine
the conductance, from which we may then infer the behavior of the commute
rate.

An advantage of formulating the average commute rate as in
Eq.~(\ref{CR-cond}) is that only sites that are connected to the starting
point $a$ contribute to ${\cal R}_a$.  Thus we can restrict the endpoint $b$
to lie in the cluster that also contains $a$.  Then averaging over many
realizations of these clusters is equivalent to averaging over $a$ in
Eq.~(\ref{CR-graph}).

\subsection{Rooted Geodesic Tree}

To generate a cluster within the random graph that contains the starting site
$a$ of the random walk, we first construct a subset of the cluster that we
term the {\it rooted geodesic tree} (RGT).  We can then build the rest of a
random graph cluster from the RGT.  The RGT is a specific subset of a random
graph cluster that: (i) spans all the sites in the cluster, and (ii) the
distance between $a$ and any site $b$ on the RGT is also the shortest
distance between these two sites in the random graph cluster.  The notion of
the RGT is inspired, in part, by the minimal spanning tree, a construction
with useful applications in network flow problems \cite{ATO,add}.

\begin{figure}[ht]
\includegraphics*[width=0.3\textwidth]{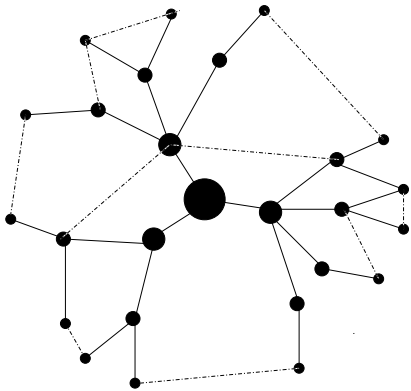}
\caption{A rooted geodesic tree (RGT).  Disks represent sites.  The largest
  black disk is the root.  Disks in successive shells have successively
  smaller radii. The solid lines are links in the RGT.  The broken lines are
  bonds that are added subsequently to generate a random graph cluster.}
\label{RGT}
\end{figure}

To construct the RGT, we start with $N$ sites and no bonds, and assign one
site (denoted by $a$) as the root; this site is defined to be at level $j=0$.
We generate the RGT as a series of successive shells centered about $a$.  The
$j^{\rm th}$ shell, denoted by ${\bf S}_j$, contains those sites that can be
reached from $a$ after exactly $j$ hops between connected neighbors.  We
define $S_j$ as the expected number of sites in the $j^{\rm th}$ shell.

Suppose that we have just generated the $j^{\rm th}$ shell.  An unassigned
site $y$ becomes part of shell ${\bf S}_{j+1}$ if a link is created that
joins $y$ to an arbitrary site $x$ in ${\bf S}_j$.  For each unassigned site,
at least one such link will be created with probability $1-(1-p)^{S_j}$.  We
then test each possible link between $y$ and the sites in ${\bf S}_j$ one by
one.  Each such link will be created with probability $p$.  When the first
such link is created, $y$ becomes an element of ${\bf S}_{j+1}$ and we then
consider the next unassigned site for potential inclusion in ${\bf S}_{j+1}$.
If none of the possible links is created, $y$ remains available for inclusion
in subsequent shells.  After all the unassigned sites have been tested, the
shell ${\bf S}_{j+1}$ is complete.  This growth process continues until
either no unassigned sites remain or if all attempts to incorporate the
available sites into the current shell fail.  In the latter case, the total
number of sites in the RGT is less than $N$.

There are two important subtleties associated with this construction
algorithm for an RGT.  First, bonds that are not examined in the initial
construction of the RGT can only exist between sites in the same or in
adjacent shells of the RGT.  A second important point is that in building the
RGT, each examined bond was tested one time only and is therefore included in
the RGT with probability $p$.

\begin{figure}[ht]
\includegraphics*[width=0.45\textwidth]{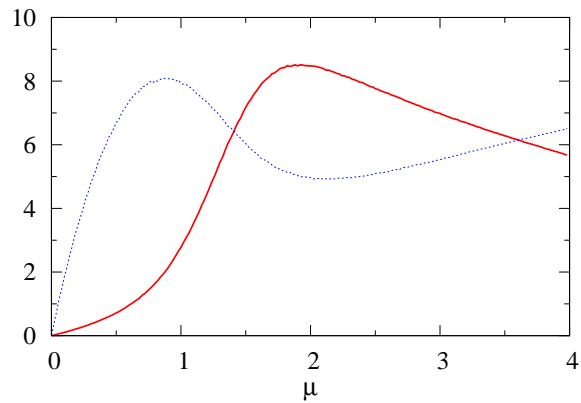}
\caption{Average RGT radius (solid curve) as a function of the average 
  degree $\mu$.  The dotted curve is the average commute rate on the same
  structure; this rate has been scaled to lie on a similar vertical range.
  Data are based on $10^5$ realizations of RGTs of 100 sites.}
\label{rgtdia}
\end{figure}

Since the number of sites in successive shells of the RGT grows exponentially
in the number of steps away from the root, the radius of the largest cluster
(giant component), is given by the criterion $\mu^L\approx \beta N$.  Here
$\beta= \beta(\mu)$ is the fraction of the initial $N$ sites that belongs to
the largest cluster.  By the definition of the RGT, the fraction of sites in
the largest cluster in the random graph and in the underlying RGT are
identical.  Thus the radius of the RGT is given by
 \begin{eqnarray}
\label{diameter}
L &\sim& \frac{\ln N}{\ln\mu} + \frac{\ln\beta}{\ln\mu}.
\end{eqnarray}
Since $\beta$ is a rapidly growing function of $\mu$ \cite{B,JLR}, the radius
of the giant component of the RGT is an increasing function of $\mu$ just
above the percolation threshold $\mu_c=1$.  This increase in radius occurs
because the RGT acquires progressively more sites with increasing $\mu$.  On
the other hand for sufficiently large $\mu$, the giant component will contain
almost all sites in the graph and the radius of this component will decrease
as $\mu$ is increased still further (Fig.~\ref{rgtdia}).  This non-monotonic
behavior of the RGT radius on $\mu$ is ultimately connected to the
non-monotonicity in the commute rate.

\subsection{Role of Loops on Graph Structure}

Given a rooted geodesic tree, it is possible to augment the tree to generate
a realization of a random graph cluster.  We merely attempt to add to the RGT
each of the bonds between sites on the cluster that were not previously
considered in the construction of the RGT itself.  Each such bond addition
attempt is carried out with probability $p$.  As a result of the fact that
each of these newly-added bonds and each bond in the RGT is present with
probability $p$, all bonds in the full random graph cluster are present with
probability $p$.  Furthermore, both the RGT and the corresponding random
graph have the same number of sites and radius for the same values of $\mu$
and $N$.  As a result of this equivalence, the RGT undergoes the same
percolation transition as the random graph itself when $\mu$ passes through
1.

\section{The Commute Rate}

We now use the connection between commute rate and conductance to understand
the non-monotonicity in the commute rate for a random walk on a random graph.
As indicated in Fig.~\ref{rgtsmall}, there are three regimes for the commute
rate: (I) an initial increase with $\mu$ for small $\mu$; (II) a decrease
over an intermediate range; and (III) an ultimate increase for large $\mu$.
For regimes I and II, the commute rates on the RGT and the random graph are
nearly identical and it is simpler to consider the commute rate on the RGT.
We then investigate how adding the links to the RGT to create a random graph
affects the commute rate.

\subsection{Commute Rate on the RGT}

For a tree graph, the resistance between two sites is simply the path length
between these two sites.  Thus the average commute rate in
Eq.~(\ref{CR-cond}) has the form
\begin{equation}
\label{CR-tree-prob}
\mathcal{R}_a = \frac{1}{V-1}\sum_{b\ne a}\frac{1}{\mathcal{D}_{a b}}
=  \frac{1}{V-1} \sum_{j=1}^{L}\frac{S_j}{j}.
\end{equation}
Here the number of links in a tree is one less than the total number of sites
$V$, and $\mathcal{D}_{ab}$ is the distance between $a$ and $b$.  Thus ${\cal
R}_a$ is the inverse moment of the distance between the root $a$ and all
other sites in the tree.  The second equality follows from the shell
structure of the RGT, where $L$ is the radius of the tree.  Thus we need only
the statistics of the shell sizes of the RGT to determine the commute rate.

\begin{figure}[ht]
\includegraphics[width=0.500\textwidth]{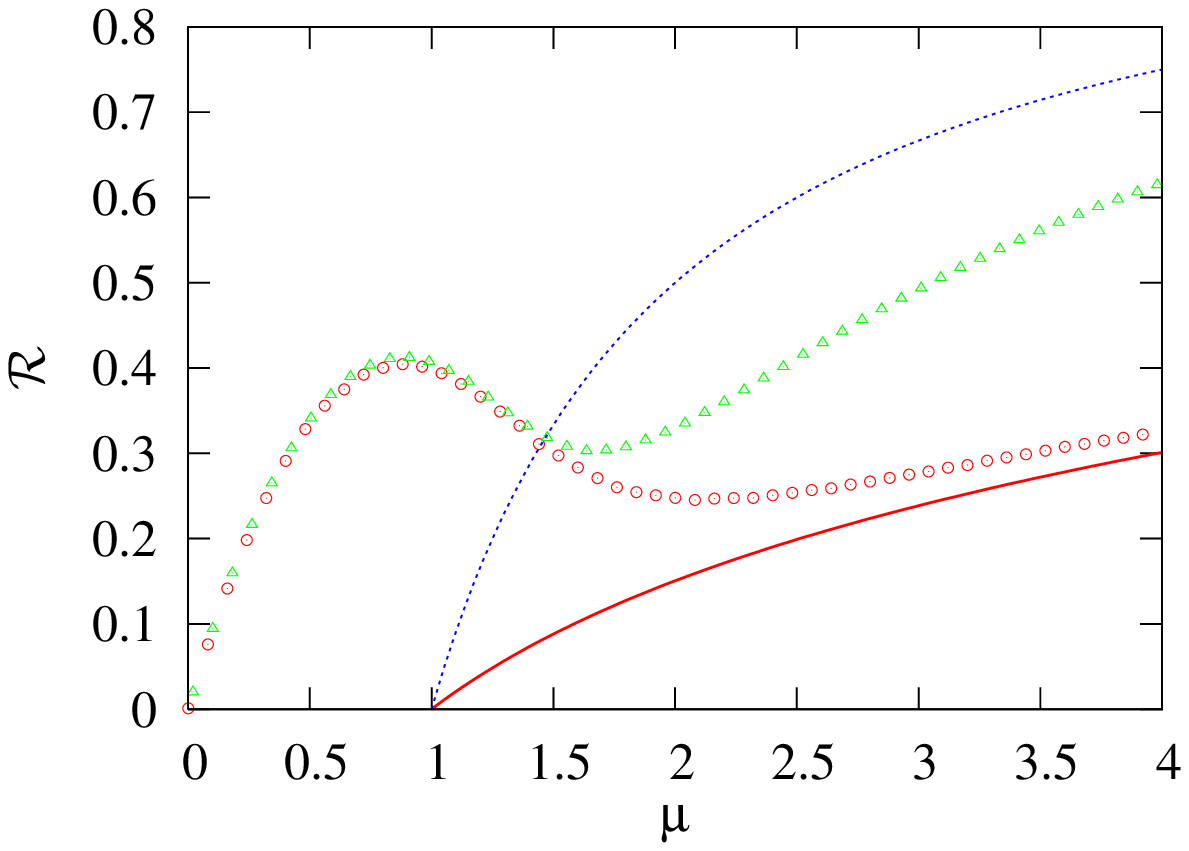}
\vskip -1.9in\hskip -0.5in
  \includegraphics[width=0.225\textwidth]{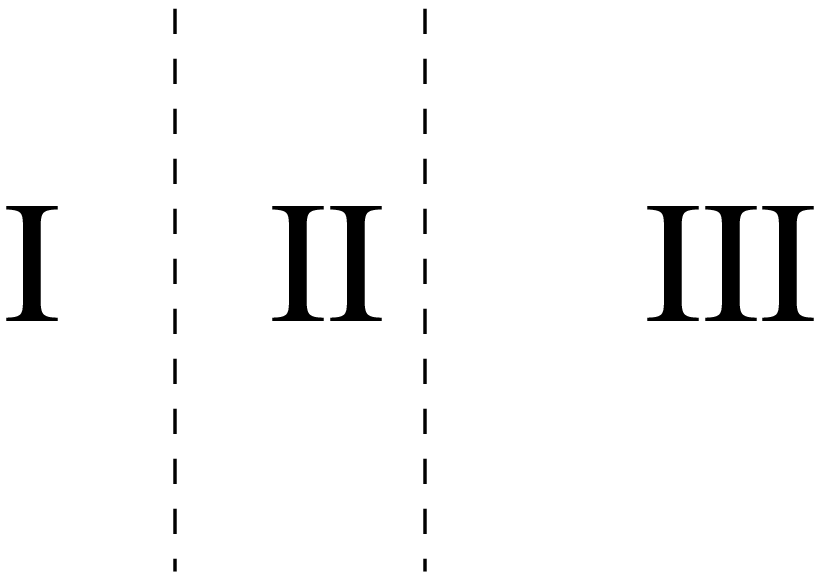}
\vskip 1.in
\caption{Commute rates on the RGT ($\circ$) and on random graphs
  ($\bigtriangleup$) for $N=100$ sites based on averages over $10^5$
  realizations.  The lower curve is our RGT prediction ${\cal R}_a=\ln\mu/\ln
  N$ for the large-$\mu$ limit.  The upper curve in the prediction for the
  random graph ${\cal R} = (\mu-1)/\mu$ (Eq.~(\ref{CR-cayley-appx})).  The
  approximate locations of regimes I, II, and III are indicated.}
\label{rgtsmall}
\end{figure}

Since each realization of the RGT is distinct, the number of sites $V$, the
radius $L$, and the shell sizes $S_j$ fluctuate from realization to
realization.  To calculate the configuration-averaged commute rate $\langle
{\mathcal R}_a\rangle$, we first use the algorithm of the previous section to
generate RGTs.  Then we solve the random walk problem on each realization and
average Eq.~(\ref{CR-tree-prob}) over realizations to determine the commute
rate (Fig.~\ref{rgtsmall}).

To understand the non-monotonicity of the commute rate for the RGT, consider
first the small-$\mu$ limit.  Because isolated sites contribute zero to the
average rate, the commute rate must initially increase with $\mu$, as small
trees begin to form.  Once most sites are no longer isolated, the radii of
typical RGTs then increase with $\mu$ due to the merging of small trees.
This increase in radius causes a decrease in the commute rate, as can be seen
by writing the rate in Eq.~(\ref{CR-tree-prob}) as
\begin{eqnarray}
\mathcal{R}_a(L) &=&  {{\displaystyle\sum_{j=1}^{L}\frac{S_j}{j}}}\,\Bigg/\,
{\displaystyle{\sum_{j=1}^{L}{S_j}}}.
\end{eqnarray}
In the analogous expression for ${\cal R}_a(L+1)$, the numerator
increases by ${S_{L+1}}/{(L+1)}$ while the denominator increases by
$S_{L+1}$.  Thus ${\cal R}_a(L)$ is a decreasing function of $L$, so that a
tree with a larger radius will have a smaller commute rate.  

As argued in Sec. V, a further increase in $\mu$ will cause the radius of the
RGT to eventually decrease with $\mu$.  Correspondingly, the commute rate
enters regime III and increases with $\mu$.  In this regime, we now use the
fact that the number of sites in successive shells of the RGT grows
exponentially in the distance from the root.  Thus the shell at radius $L$
contains almost all of the sites of the RGT.  As a naive approximation, we
then replace the sum in Eq.~(\ref{CR-tree-prob}) by the last term to give, in
the limit of large $\mu$,
\begin{eqnarray*}
(\mathcal{R}_a)_{\rm RGT} \approx \frac{1}{V}\,\, \frac{V}{L}
\sim \frac{\ln\mu}{\ln N}.
\label{CR-RGT-asymp}
\end{eqnarray*}
This result agrees extremely well with numerical results for the commute rate
on RGTs, as shown in Fig.~\ref{rgtsmall}.

\subsection{Role of Loops on Commute Rate}

\begin{figure}[ht]
  \includegraphics*[width=0.35\textwidth]{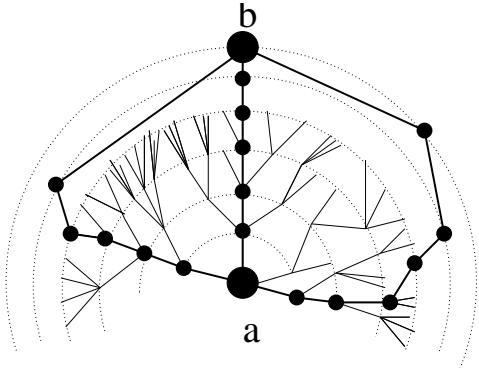}
\caption{Schematic representation of the random graph.  The included RGT is
  also shown.  Loops typically arise at a distance $L = {\ln N}/{\ln\mu}$
  from the root.  A site in this last shell will typically have $\mu$
  independent paths to the root.  }
\label{RGloops}
\end{figure}

We now investigate how adding loops to the RGT to build a random graph
affects the behavior of the mean commute rate.  Starting with a realization
of an RGT we generate a cluster of the random graph by adding missing bonds,
following the procedure discussed in Sec.~V.  The addition of these bonds
will create loops that provide alternative paths between the root site and
the endpoints of a random walk (Fig.~\ref{RGloops}).  The ostensible effect
of these additional paths is to increase the commute rate between the root
and any endpoint.

\begin{figure}
\includegraphics*[width=0.3\textwidth]{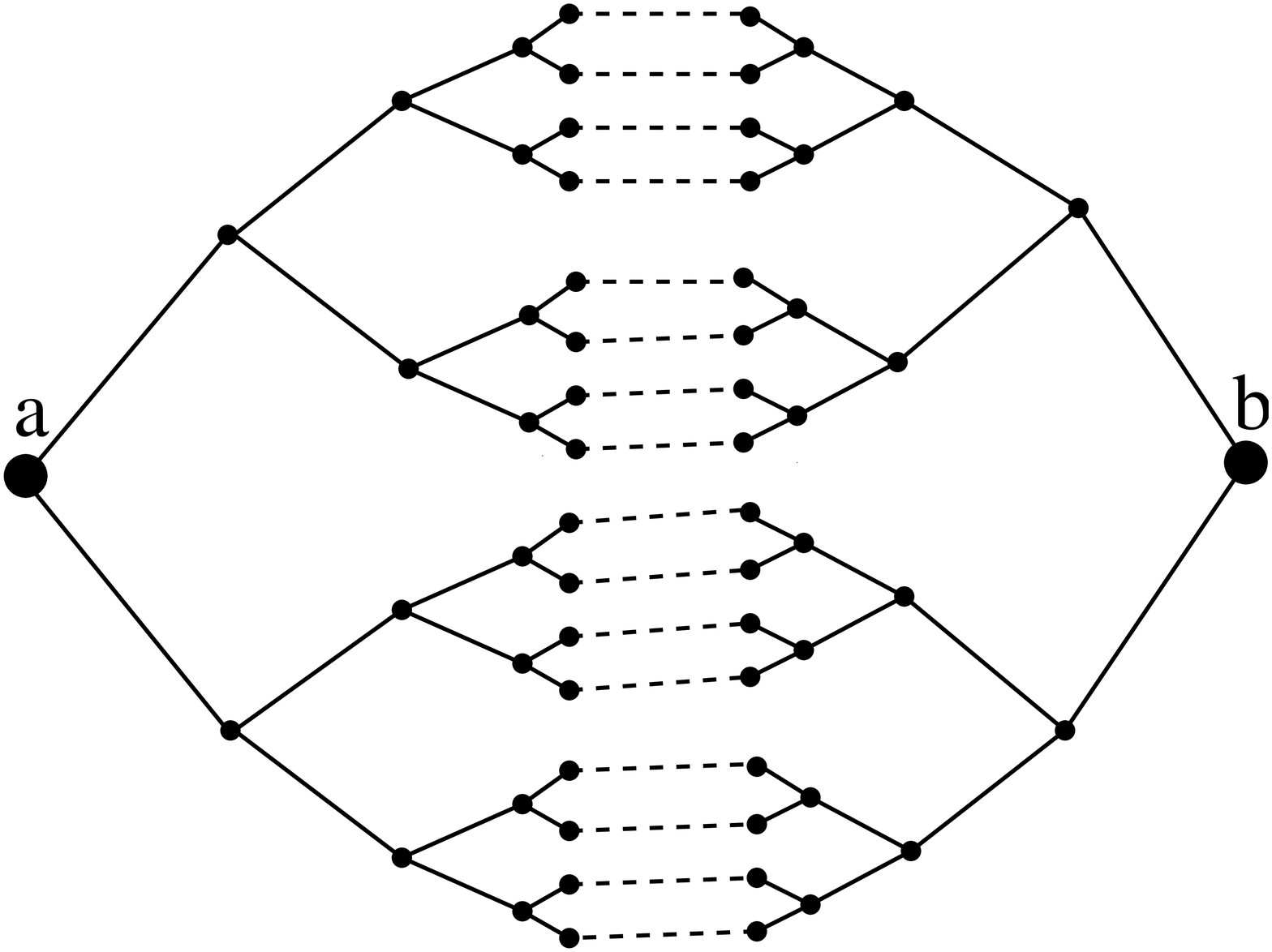}
\caption{Schematic random graph structure to calculate the conductivity
between two sites $a$ and $b$ for mean degree $\mu>1$.  An RGT is grown
around both $a$ and $b$.  The two RGTs meet at a distance $O(\ln N/\ln\mu)$
from each of $a$ and $b$.  Broken lines are links between sites in the
outermost shells of the two respective RGTs. }
\label{twoRGTs}
\end{figure}

To estimate the two-point conductance for a random graph for general
$\mu>{\rm ln}~N$, we start with the picture that the graph consists of two
RGTs, one emanating from $a$ and the other from $b$ (Fig.~\ref{twoRGTs}).
For a graph of $N$ sites, the radius of each tree is of order $L\sim
\ln(N/2)/\ln\mu$.  We argue that these two trees tend to join only at the
outermost shell because this is where most of the sites in the trees are
located.  We further assume that, in the equivalent resistor network, all
sites at the same distance from the root are at the same potential.  Thus the
conductance of the two joining RGTs is simply one-half of the conductance
between the root and the last shell of a single RGT.

For this last step, we approximate the RGT by an infinite Cayley tree with
branching ratio $\mu$.  The resistance between the $k$-th and the $(k+1)^{\rm
st}$ shell in this tree is $\mu^{-(k+1)}$, since the links between the two
shells are in parallel.  Because the shells are in series, the resistance
from the center to infinity is simply the geometric sum, $\sum_{k=0}^{\infty}
\mu^{-(k+1)} = \frac{1}{\mu-1}$.  Thus the conductance between $a$ and $b$ is
$G_{a b} = (\mu-1)/2$.  Substituting this result in
Eq.~(\ref{CR-DenseGraph}), then gives the commute rate
\begin{equation}
\label{CR-cayley-appx}
{\cal R}= \frac{\mu-1}{\mu}.
\end{equation}
This result converges to 1 as $\mu\to\infty$, in agreement with the effective
medium approach in Sec II as well as our simulation results.  Closer to the
percolation threshold, however, Eq.~(\ref{CR-cayley-appx}) and simulation
results quantitatively disagree because our naive picture for the structure
of the random graph no longer applies.

Thus we observe that the eventual increase in the commute rate (regime III)
stems from the combined effect of the decrease in the radius of the
underlying RGT embedded within a random graph cluster and the emergence of
loops that join two RGTs in the random graph.

\section{Conclusion}

We studied a basic first-passage characteristic of random walks on random
graphs that is related to the time for a walk to travel between two arbitrary
points on a graph.  We first constructed an effective medium theory and a
small dilution approximation for this mean transit time.  The former approach
predicted that the mean transit time, and also all positive integer moments
of the transit time, are independent of the bond concentration $p$ for $p$
greater than the connectivity threshold $p_1=\ln N/N$.  The small dilution
approximation also predicts a slow dependence of the transit time on $p$ near
$p=1$.  Our numerical simulation results are in qualitative accord with a
transit time that is slowly varying in $p$ for $p>p_1$.

Below the connectivity threshold, the transit time is not well defined
because the mean time for a random walk to hop between sites on different
components of a disconnected graph is infinite.  To avoid this pathology, we
studied the inverse of the commute time, namely, the commute rate.  We
developed a simple heuristic picture for the behavior of the commute rate
that relied on first identifying an embedded rooted geodesic tree (RGT)
within an arbitrary random graph cluster.  For the RGT, it is simple to
compute the commute rate in terms of a geometric picture for the tree and
thus argue that this rate is a non-monotonic function of $p$ in the critical
regime.

We then presented a simple physical picture for the influence of loops on the
behavior of the commute rate.  Qualitatively, the dependence of the radius of
the underlying RGT on the bond concentration explains the behavior of the
commute rate close to the percolation threshold.  For larger $\mu$ loops
become an important factor and are ultimately responsible for the
non-monotonic dependence of the commute rate on $p$.  While our arguments
were heuristic and the approximations made are uncontrolled, they provide an
intuitive picture for the structure of random graphs and also provide
qualitative and satisfying agreement with simulation results for the commute
rate.

\acknowledgments{ We thank Paul Krapivsky and Federico Vazquez for helpful
  suggestions.  VS and SR gratefully acknowledge financial support from NSF
  grants DMR0227670 (at BU) and DOE grant W-7405-ENG-36 (at LANL).  DbA
  similarly acknowledges NSF grant PHY0140094 (DbA) for financial support.  }

\end{document}